\documentstyle[aps,twocolumn]{revtex}

\begin{document}

\title{Spontaneously ordered motion of self-propelled
particles} \author{
Andr\'as Czir\'ok$^{1,2}$,
H.~Eugene~Stanley$^1$ and
Tam\'as~Vicsek$^2$}

\address{
$^1$ Center for Polymer Studies and Department of Physics,
Boston University, Boston, MA 02215\\
$^2$ Department of Atomic Physics,
E\"otv\"os University, Budapest, Puskin u. 5-7, 1088 Hungary \\
}

\maketitle

\begin{abstract}

We study a biologically inspired, inherently non-equilibrium model
consisting of self-propelled particles.  In the model, particles move on
a plane with a velocity of constant magnitude; they locally interact with
their neighbors by choosing at each time step a velocity direction equal to
 the average
direction of their neighbors. Thus, in the limit of vanishing velocities 
the model becomes analogous to a Monte-Carlo realization of the
classical XY ferromagnet. We show by
large-scale numerical simulations that, unlike in the equilibrium 
XY model, a long-range ordered
phase characterized by non-vanishing net flow $\phi$ emerges in this system
in a phase space
domain bordered by a critical line along which the fluctuations of the
order parameter diverge.
The corresponding
phase diagram  as a function of two parameters, the
amplitude of noise $\eta$ and the average density of the particles
$\varrho$ is calculated and is found to have the form $\eta_c(\varrho)\sim
\varrho^{1/2}$. We also find that $\phi$ 
scales as a function of the external bias $h$ (field or ``wind'')
according to a power law $\phi\sim h^{0.9}$.
  In the ordered phase the system shows
long-range correlated fluctuations and $1/f$ noise.

\end{abstract}

\section{introduction}

Recently there has been an increasing interest in the studies of
far-from-equilibrium systems typical in our natural and social
environment. Concepts originated from the physics of phase transitions
in equilibrium systems \cite{HES71} such as collective behavior, scale
invariance and renormalization have been shown to be useful in the
understanding of various non-equilibrium systems as well.  Simple
algorithmic models have been helpful in the extraction of the basic
properties of various far-from-equilibrium phenomena, like diffusion
limited growth \cite{DLA}, self-organized criticality\cite{SOC} or
surface roughening \cite{surface}. Motion and related transport
phenomena represent a further characteristic aspect of non-equilibrium
processes. Indeed, the transport in various driven systems, such as
traffic models\cite{traffic}, molecular motors\cite{MM} and other
self-propelled systems \cite{VCBCS95,CV95,TT,DHT,Hem95,brasil} have been the
subject of recent studies.

Self-propulsion is an essential feature of most living systems.
Moreover, the motion of the organisms is usually controlled not only by
some external fields, but also by interactions with other organisms in
their neighborhood. In Ref.\cite{VCBCS95}, a simple model was introduced
capturing these features with a view toward modeling the collective
motion of large groups of organisms \cite{Reynolds87,OBrien89,DG89,HW90} such as
schools of fish, herds of quadrupeds, flocks of birds, or groups of
migrating bacteria\cite{CBCV,AH91,BTSA94,BCSTCV94,BCSTCV95}.  The aim of
this paper is to further investigate the various interesting phenomena
exhibited by this novel non-equilibrium model.

The model consists of particles moving on a plane and characterized by
their (off-lattice) location $\vec{x}_i$ and velocity $\vec{v}_i$
pointing in the direction $\vartheta_i = \Theta (\vec{v}_i)$, where the
function $\Theta$ gives the angle between its argument vector and a
selected direction (e.g., horizontal coordinate axis).  The magnitude of
the velocity is fixed to $v_0$ to account for the {\it self-propelled}
nature of the particles.  A simple local interaction is defined in the
model: at each time step a given particle assumes the average direction
of motion of the particles in its local neighborhood $S(i)$ with some
uncertainty, as described by
\begin{equation}
\vartheta_i (t+\Delta t) = \langle \vartheta(t) \rangle_{S(i)} + \xi,
\label{EOM}
\end{equation}
where the noise $\xi$ is a random variable with a uniform distribution
in the interval $[-\eta/2,\eta/2]$ and the local average direction of
motion $\langle \vartheta \rangle_{S(i)}$ is defined as
\begin{equation}
\langle \vartheta \rangle_{S(i)} =
\Theta\left(\sum_{j\atop \vec{x}_j \in S(i)} \vec{v}_j \right).
\label{avg}
\end{equation}
The local surrounding of the $i$th particle $S(i)$ will be specified
in Sec. II. 
The locations of the particles are updated in each time step as
\begin{equation}
\vec{x}_i(t+\Delta t) = \vec{x}_i(t) + \vec{v}_i(t)\Delta t.
\label{update}
\end{equation}

This model is a transport related, non-equilibrium analog of the {\it
ferromagnetic} models, with the important difference that it is
inherently {\it dynamic}: the elementary event is the motion of a
particle at each time step and a change in the direction of motion.  The
analogy is as follows: the Hamiltonian tending to align the spins in the
same direction in the case of equilibrium ferromagnets is replaced by
the rule of aligning the direction of motion of particles.  The
amplitude of the random perturbations is in analogy with the temperature
\cite{VCBCS95}. Indeed, if $v_0 = 0$, the model is similar to the
Monte-Carlo simulations of diluted $XY$ ferromagnets \cite{XY}.

The reported long-range order in the above\cite{VCBCS95} and the closely
related models\cite{DHT,Hem95} is surprising, because in the case of
equilibrium systems possessing continuous rotational symmetry the
ordered phase is destroyed at finite temperatures\cite{MW66}.  A recent
dynamic renormalization group treatment of the problem\cite{TT} by Tu
and Toner has also led to the conclusion of the existence of an ordered
phase in two dimensions. Thus, the question of how the ordered phase
emerges due to the non-equilibrium nature of the model is of
considerable theoretical interest as well.  In Section II we study the
kinetic phase transition leading to the symmetry-broken state, which is
characterized in Section III.  In Section IV. we investigate the effect
of an external field applied to the system.

\section{Kinetic phase transition}

We studied the behavior of the model defined through Eqs.
(\ref{EOM})-(\ref{update}) by performing large-scale Monte-Carlo
simulations as a function of two control parameters: the density of
particles $\varrho$ and amplitude of the noise $\eta$.  We applied
random initial conditions and periodic boundary conditions. The
calculations were performed on a {\it Connection Machine 5\/} parallel
computer, with typically $N=10^4-10^5$ particles.

The interaction range of $S(i)$ was defined in two different ways: (i)
as a circle of radius $R$, or (ii) by considering a square lattice on
the plane built up from lattice cells of length $R$, and assuming that a
given particle interacts with all the particles located in the same
lattice cell and in the eight neighboring cells (see Fig.~1). The
existence of the long-range order, and the critical exponents, turned
out to be {\it robust} against changing these details [(i) or (ii)] of
the interaction.  Here we present results obtained by definition (ii),
as this latter choice of $S(i)$ increases the speed of the simulations
by a considerable amount.

A natural dimensionless parameter is $C\equiv v_0\Delta t/R$, and the
behavior of the model on $v_0$, $\Delta t$ and $R$ depends only through
their combination given in this expression for $C$. Thus, we can work
with $\Delta t = 1$ and $R=1$.  The results presented in the following
were obtained with $v_0=0.1$, the role of this velocity is discussed in
Sec. V.

For the statistical characterization of the model, a well-suited order
parameter is the magnitude of the average momentum of the system:
\begin{equation}
\phi\equiv{1\over N}\left\vert \sum_j \vec{v}_j \right\vert.
\end{equation}
This measure of the net flow is non-zero in the ordered phase, and
vanishes (for an infinite system) in the disordered phase.

We start the simulations from a disordered system (random positions and
orientations), thus $\phi(t=0)\approx 0$.  After some relaxation time a
steady state emerges indicated, e.g., by the convergence of the
cumulative average $(1/\tau) \int^\tau_0 \phi(t)dt$. Here we focus on
the statistical properties of the steady state only, and do not deal
with the entire relaxation process.  In the vicinity of the critical
regime to reach the stationary behavior takes more than $10^4$
Monte-Carlo steps for a typical simulation with $10^5$ particles in a
$100\times100$ system. In such cases we run the simulations for $\approx
10^5$ time steps, which takes about $4$ hours CPU time on the $CM5$.

The stationary values of $\phi$ that we obtained as a time average are 
plotted in Fig.~2a
vs $\eta$ for $\varrho = 2$ and various system sizes.  In agreement with
Ref. \cite{VCBCS95}, for weak noise the model displays long-range
ordered motion (up to the actual system size $L$), that disappears in a
continuous manner by increasing $\eta$.

As $L \rightarrow \infty$, the numerical results indicate the presence
of a kinetic phase transition described by
\begin{equation}
\phi(\eta)\sim \cases{
         \Bigl({\eta_c(\varrho,L) - \eta\over \eta_c(\varrho,L)}\Bigr)^\beta
                & for $\eta<\eta_c(\varrho,L)$ \cr
        0  & for $\eta>\eta_c(\varrho,L)$ \cr
    },
\label{scale}
\end{equation}
where $\eta_c(\varrho,L)$ is the critical noise amplitude that separates
the ordered and disordered phases.  For a given set of $\phi(\eta)$, the
exponent $\beta$ and $\eta_c(\varrho,L)$ is determined by selecting the
values providing the best fit to the {\it Ansatz} (\ref{scale}), i.e.,
yielding the maximal scaling regime.  The numerical results are
consistent with (\ref{scale}), since the scaling regime is increased for
larger $N$ (Fig.~2b) and the estimated values of $\eta_c(\varrho,L)$
converge to a non-zero $\eta_c(\varrho,\infty)$ value as
\begin{equation}
\eta_c(\varrho,L)-\eta_c(\varrho,\infty) \sim N^{-\zeta},
\end{equation}
where $\zeta$ is approximately equal to $0.25$ (Fig.~2c).  This
calculation yields (for $\varrho = 2$) $\beta=0.42\pm0.03$, which is
definitely different from the the mean-field value $1/2$, and consistent
with the value reported in \cite{VCBCS95} obtained for smaller systems
using definition (i) for $S(i)$.

Next we discuss the role of density. In Fig.~3a, $\phi(\eta)$ is plotted
for $L=100$ and various values of $\varrho$. One can observe that the
long-range ordered phase is present for any $\varrho$, but for a fixed
value of $\eta$, $\phi$ vanishes with decreasing $\varrho$.  These
$\phi(\eta)$ functions parameterized by various $\varrho$ collapse to a
``universal'' function $\tilde{\phi}(x)$ by rescaling $\eta$ with
$\eta_c(\varrho)$,
\begin{equation}
\phi(\eta,\varrho) = \tilde{\phi}\bigl(\eta/\eta_c(\varrho)\bigr),
\label{v}
\end{equation}
where $\tilde{\phi}(x)\sim (1-x)^\beta$ for $x<1$, and $\tilde{\phi}(x)
\approx 0 $ for $x>1$, and $\eta_c(\varrho)$ is determined as the value
which minimizes
\begin{equation}
\Delta(z)=\left[\int_0^A dx\left(
\phi(xz,\varrho)-\tilde{\phi}(x)\right)\right]^2.
\end{equation}
Here the $A<1$ cutoff is chosen to exclude the noisy and rounded (due to
finite-size effects) region around $\eta_c$. In our calculations we used
the value $A=0.9$, and we determined $\tilde{\phi}(x)$ in a
self-consistent manner by averaging over the already rescaled
$\phi(\eta/\eta_c(\varrho))$ functions. We show the result of the data
collapse in Fig.~3b.

This procedure (together with the finite-size analysis for a given
$\varrho$) also yields the position of the ``critical line''
$\eta_c(\varrho)$ in the $\eta-\varrho$ parameter space.  According to
our numerical results,
\begin{equation}
\eta_c(\varrho)\sim\varrho^\kappa
\label{kappa}
\end{equation}
holds with $\kappa = 0.45 \pm 0.05$ (see Fig.~3c). Apart from the
numerical uncertainties, $\kappa = 1/2$, in agreement with recent
theoretical results \cite{BC}.

The critical line (\ref{kappa}) is qualitatively different from that of
the diluted ferromagnets, since here the critical density at
$\eta\rightarrow0$ (corresponding to the percolation threshold for
diluted ferromagnets, see, e.g., \cite{XY}) is vanishing,
\begin{equation}
\lim_{\eta\rightarrow 0}\varrho_c(\eta) = 0.
\end{equation}
To see this, let us imagine a system of size $L$ consisting of two
particles only.  Due to the finite size of $S(i)$, for almost all
initial conditions the trajectories of these particles will get close
enough to each other to establish interaction.  As the noise is
negligible in (\ref{EOM}), the cluster formed by the particles will not
break apart, resulting in $\phi\approx1$ for any finite $\varrho$ as
$\eta\rightarrow 0$.

The behavior of the model in the $\varrho\rightarrow\infty$ limit is
still not clear. By definition $\eta<2\pi$ holds, so (\ref{kappa})
obviously cannot describe the system in this limit. Thus
$\eta_c(\varrho)$ either approaches $2\pi$ or a non-trivial
$\eta_c(\infty)<2\pi$ value.

Finally we note that Eq.(\ref{v}) also implies that the exponent
$\beta'$, defined as $\phi\sim(\varrho-\varrho_c)^{\beta'}$ for
$\varrho>\varrho_c$ (see Ref. \cite{VCBCS95}), must be equal to $\beta$,
since
\begin{equation}
\phi\left(\eta,\varrho_c(\eta)+\epsilon\right)=
\tilde{\phi}\left({\eta\over\eta_c[\varrho_c(\eta)+\epsilon]}\right)\approx
\tilde{\phi}\left(1-{1\over\eta}{\partial\eta_c\over\partial\varrho}
\epsilon\right)\sim\epsilon^\beta,
\end{equation}
where $\eta_c(\varrho)$ denotes the inverse function of
$\varrho_c(\eta)$ as $\eta\equiv\eta_c[\varrho_c(\eta)]$. Indeed, the
results of the simulations performed with $\eta=const$ and various $L$
and $\rho$ yield $\beta'=0.4 \pm 0.05$, which is consistent with
$\beta=\beta'$. In this case the larger uncertainty is due to the
increased noise at low densities.

\section{Fluctuations}

As a further analogy with equilibrium phase transitions, we note that
the fluctuations of the order parameter also increase on approaching the
critical line. To study this, we calculate for various control
parameters the standard deviation of the total momentum, defined as
$\sigma^2\equiv\langle\phi^2\rangle-\langle\phi\rangle^2$, where the
averages are taken over the stationary data set obtained from the
simulations. In Fig.~4a we plot $\sigma$ vs the rescaled noise amplitude
$x\equiv\eta/\eta_c(\varrho)$ for various densities and $L=100$.  The
tails of the curves are symmetric, and decay as power-laws with an
exponent $\gamma$ close to $2$ (see Fig.~4b)
\begin{equation}
\sigma(x)\sim \vert 1-x \vert^{-\gamma}.
\end{equation}
In Sec. IV we compare this result to direct susceptibility measurements,
when an external field is also applied.

We studied the time correlations of the fluctuations by calculating the
expected value of the rms deviation in a time interval $\Delta t$
\begin{equation}
w^2_\phi(\Delta t)=\langle \langle \phi^2 \rangle_{\Delta t} -
\langle \phi \rangle^2_{\Delta t} \rangle,
\label{w}
\end{equation}
where the internal and external brackets denote averages calculated over
a time interval $[t, t+\Delta t]$ and the entire stationary data set,
respectively. A close relation between $w_\phi(\Delta t)$ and the
power-spectrum $S_\phi(\omega)$ of the order parameter $\phi(t)$ can be
derived from the self-affine properties of the signal\cite{VOSS}:
$w_\phi(\Delta t)\sim(\Delta t)^\alpha$ is equivalent to
$S_\phi(\omega)\sim\omega^{-\lambda}$, and for the exponents $\lambda =
1+2\alpha$ holds (for $3>\lambda>1$).

The numerically obtained results (Fig.~5a) show that at the critical
line the fluctuations of the order parameter are characterized by the
correlation exponent $\alpha = 0.6 \pm 0.05$ up to a characteristic
correlation time $\tau$. This behavior means that in the steady state
the `condensation' and `evaporation' processes (when particles join or
leave the dominant cluster, respectively) are correlated \cite{NoteI}.
For $t>\tau$ the the system shows even stronger correlations, as in this
regime $w_\phi(\Delta t) \sim \log\Delta t$ (see Fig.~5b). This behavior
is probably related to the periodic boundary conditions, as $\tau$ is
comparable to $L/v_0$, the time needed for a particle to cross the
entire system.

\section{External field}

The presence of the long-range correlated phase in two dimensions
(i.e. the breakdown of the Mermin-Wagner theorem \cite{MW66}) is
a striking consequence of the non-equilibrium nature of the $XY$ model. As
the fluctuation-response theorem $kT\chi = N\sigma^2$ must hold for
Hamiltonian systems only, it is interesting to check its applicability
when the equation of motion cannot be derived from a Hamiltonian.

A natural way to introduce an external field in the model is by changing
rule (\ref{avg}) for
\begin{equation}
\langle \vartheta \rangle_{S(i)} =
\Theta\left( \sum_{j\atop \vec{x}_j \in S(i)} \vec{v}_j + h\vec{e} \right).
\label{avg2}
\end{equation}
where $\vec{e}$ is an arbitrary unit vector and the parameter $h$
controls the strength of the perturbation.

In the numerical studies we set the density of the particles to $\varrho
= 2$, and study $\phi$ as a function of $\eta$ and $h$.  Typical results
for systems of $L=100$ are plotted in Fig.~6.  The curves parameterized
by various $\eta$ intersect the $h=0$ axis at the same values found in
Sec. II, so for $h=0$ the original model is recovered. In general,
$\phi$ is increased for increasing $h$, and the results can be
summarized by means of critical exponents $\delta$, $\gamma_*$,
$\gamma_*'$ and susceptibility $\chi(\eta)$ defined in a manner similar
to the case of classical magnets:
\begin{equation}
\chi(\eta) = \lim_{h\rightarrow 0} {\phi(\eta,h)-\phi(\eta,h=0)\over h}, 
\end{equation}
\begin{equation}
\phi \sim h^{1/\delta}\hbox{~~~~~~for~~~~} \eta>\eta_c,
\end {equation}
and
\begin{equation}
\chi(\eta) \sim \cases {
        \Bigl( {\eta - \eta_c \over \eta_c} \Bigr)^{-\gamma_*}
          & for $\eta>\eta_c$ \cr
        \Bigl( {\eta_c - \eta \over \eta_c} \Bigr)^{-\gamma_*'}
          & for $\eta<\eta_c$ \cr }
\label{def}
\end{equation}
We distinguish the critical exponents $\gamma_*$ and $\gamma_*'$ from
the exponent $\gamma$ defined by the singularity of $\sigma(\eta)$,
since in this case $\gamma\neq \gamma_* \neq \gamma_*'$ (as will be
demonstrated later).

To determine $\delta$, we plot the obtained $\phi(\eta)$ curves
on a double logarithmic plot (see Fig.~7a). We see that for
large values of $\phi$ the system saturates
($\phi>\phi_{\mbox{\scriptsize sat}}\approx 0.1$), while for 
low values of $\phi$ the finite-size noise dominates
($\phi<\phi_{\mbox{\scriptsize noise}}\approx 0.03$, for
$L=100$).  Thus, we expect that the relation
\begin{equation}
\phi(\eta,h)=\chi(\eta)h^{1/\delta}
\label{delta}
\end{equation}
holds for $\phi_*\in [\phi_{\mbox{\scriptsize noise}},
\phi_{\mbox{\scriptsize sat}}]$ only. As the boundaries 
of this interval do not depend on $\eta$ until $\eta>\eta_c$,
Eq. (\ref{delta}) can be verified by collapsing the rescaled
$\phi_*(\eta,h)/\chi(\eta)$ data points onto a single power-law 
(Fig.~7b). Similarly to the determination of $\eta_c(\varrho)$,
$\chi(\eta)$ and $\delta$ can be calculated in a self-consistent
manner. This procedure yields $\delta = 1.1 \approx 1$ and
$\chi(\eta)$ decaying with $\gamma_* \approx 4$. Note, that
$\gamma_*'$ cannot be obtained by this method as in the ordered
phase $\phi>\phi_{\mbox{\scriptsize sat}}$. 

To check the stability of these results, we also performed measurements
on larger systems, where $\phi_{\mbox{\scriptsize noise}}$ is reduced. Indeed, the scaling
regime for (\ref{delta}) increases and the exponent is consistent with
our former estimate (see Fig.~7c)

We can also calculate $\chi(\eta)$ applying its definition (\ref{def})
by sampling a series of $\chi(\eta)_i\equiv [\phi(\eta,h_i) -
\phi(\eta,h_i') ] /(h_i-h_i')$ for various $h_i, h_i' \ll 1$, and (as
$\delta \approx 1$) assuming $\chi(\eta) = \langle \chi_i(\eta)
\rangle_i$.  The result is shown on Fig.~8. The $\eta>\eta_c$ tail is
consistent with our former estimate from applying (\ref{delta}), while
for $\eta<\eta_c$ $\chi(\eta)$ decays with $\gamma_*'\approx 1$.

\section{Conclusions}

We have demonstrated that this far-from-equilibrium system of
self-propelled particles can be described using the framework of
classical critical phenomena, but shows surprising new features when
compared to the analogous equilibrium systems.  The velocity $v_0$
provides a control parameter which switches between an equilibrium $XY$
type model ($v_0=0$), and another universality class of dynamic,
non-equilibrium models characterized by a non-vanishing value of $v_0$.

Indeed, for $v_0=0$ we can observe Kosterlitz-Thouless vortices in the
system, which turned out to be unstable for any nonzero $v_0$ we
investigated: in Fig.~9 we plot snapshots of the system. These pictures
demonstrate how the vortices disappear and give way to long-range order.
We performed control simulations with various $v_0$ in the range $[0.01,
0.3]$, and the results presented seem to be robust against changing the
value of $v_0$ in this range.

\acknowledgements

We have benefited from discussions with E. Ben-Jacob, Z. Csah\'ok,
S. T. Harrington and H. Makse. This work was supported by the
US-Hungarian Joint Fund Contract No. 352. and by the Hungarian Research
Foundation grant No. T4439. A. Czir\'ok is grateful to A.-L. Barab\'asi
for his kind hospitality during his visit at University of Notre Dame.

\begin{figure}
\caption{Scematic illustration of the model. The particles 
move off-lattice on a plane and interact with other particles
located in the local surrounding,  which can be either a
circle or 9 neighboring cells in an underlying
lattice. We plot these interaction areas for particle $A$ with a
solid and dashed line, respectively.
}
\end{figure}

\begin{figure}
\caption{
(a) The average momentum of the system in the steady state vs the noise
amplitude $\eta$ for $\varrho = 2$ and four different system sizes
[($\diamond$) $N=800$, $L=20$; ($+$) $N=3200$, $L=40$; ($\Box$) $N=20000$,
$L=100$ and ($\times$) $N=10^5$, $L=223$]. (b) The order present at small
$\eta$ disappears in a continuous manner reminiscent of second order
phase transitions: $\phi\sim [(\eta_c(L)-\eta)/\eta_c(L)]^{\beta} \equiv
(\Delta\eta/\eta_c)^{\beta}$, with $\beta=0.42$, different from the
mean-field value $1/2$ (solid line). (c) The estimated $\eta_c(L)$ values
converge to a non-zero $\eta_c(\infty)$ limit as $L^{-1/2}$, indicating
the presence of the ordered phase even as $N\rightarrow\infty$. The data
represent time averages of long ($>10^5$ MCS) simulations.
}
\end{figure}

\begin{figure}
\caption{
(a) The average momentum of the system in the steady state vs $\eta$ for
$L=100$ and three different densities [($\Box$) $\varrho = 4$,
($\Diamond$) $\varrho = 2$ and ($+$) $\varrho = 0.5$]. (b) The
$\phi(\eta)$ functions parameterized by various $\varrho$ can be
collapsed onto a single curve $\tilde{\phi}[\eta/\eta_c(\varrho)] \equiv
\phi(\eta,\varrho)$.  (c) The critical line in the $\eta-\varrho$ phase
space is a power-law in the examined regime:
$\eta_c(\varrho)\sim\varrho^{\kappa}$ with $\kappa\approx 0.45$ (solid
line) for a system of size $L=100$.  }
\end{figure}

\begin{figure}
\caption{
(a) The rms deviation of the order parameter $\sigma$ in the steady
state, for $L=100$, and two independent data sets ($\Diamond,+$), each
averaged over runs with $\varrho=0.3, 0.6, 0.9, 1.2, 1.5$ and $2$ .  (b)
The divergence is symmetric and its tail decays as $\sigma(x)\sim \vert
1-x\vert^{-2}$ (solid line), where $x$ denotes the rescaled noise
amplitude $\eta/\eta_c(\varrho)$. }
\end{figure}

\begin{figure}
\caption{The expected value of the standard deviation of the order
parameter in a time interval of length $\Delta t$ on double logarithmic
(a) and log-linear (b) plots. For $\Delta t<\tau \approx L/v_0$ the
system shows long-range correlations characterized by $w_\phi(\Delta
t)\sim\Delta t^\alpha$ where $\alpha \approx 0.6$.  For comparison, the
dotted line shows the uncorrelated case ($\alpha = 1/2$).  For $\Delta t
> \tau$ the correlations are even stronger: $w_\phi(\Delta t) \sim
\log(\Delta t)$. The data displayed is an average over two independent
runs, each was performed at $\eta=2.73$, $\varrho=2$ and $N=10^5$. The
duration of the simulation was 70~000 Monte-Carlo steps.}
\end{figure}

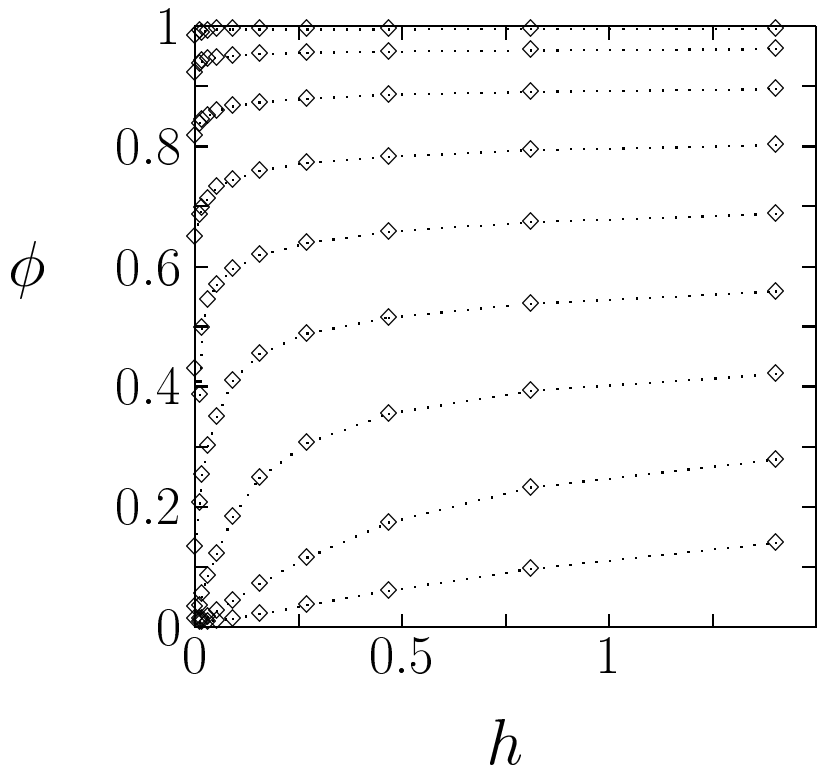
\begin{figure}
\caption{
$\phi$ vs the amplitude of the applied external field $h$, for
$\varrho=2$, $L=100$ and $\eta=0.3, 0.9, 1.5, 2.1, 2.7, 3.3, 3.5$ and
$4.5$ (from top to bottom).  }
\end{figure}

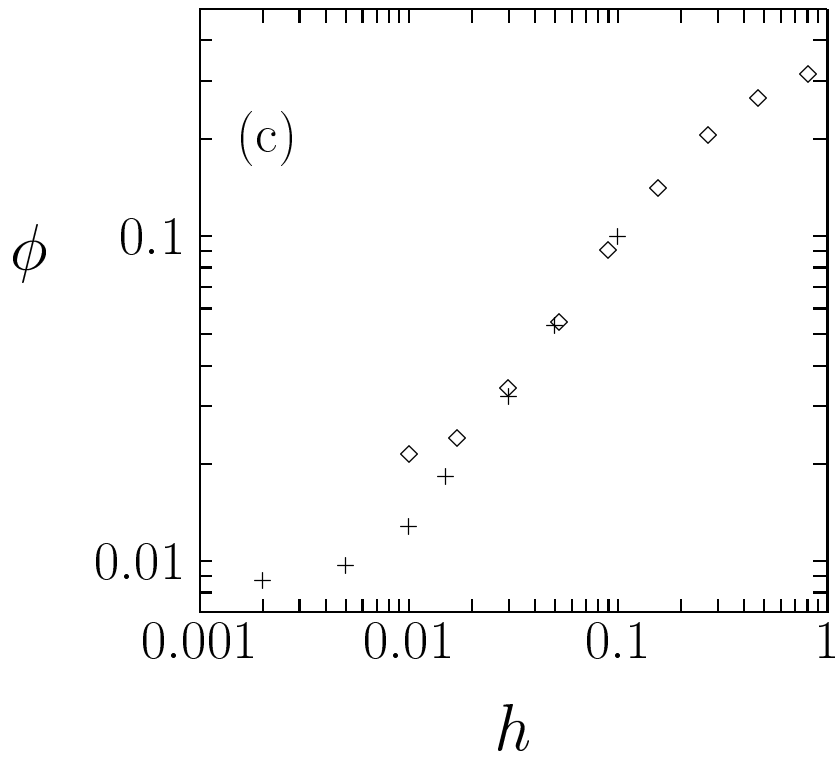
\begin{figure}
\caption{
(a) The $\phi(h)$ functions for $\varrho=2$, $L=100$ and
$\eta=3.9,4.2,4.5,4.8, 5.2$ (from top to bottom) on a double-logarithmic
plot.  $\phi\sim h^{1/\delta}$ holds for a limited range of $\phi$ only,
as for $\phi>0.1$ the system saturates, and for $\phi<0.02$ the
finite-size noise dominates. (b) For $0.02<\phi<0.1$ the curves for
various $\eta$ can be collapsed onto a single power-law with an exponent
$1/\delta \approx 0.9$.  (c) For larger systems the finite-size noise is
reduced and the scaling regime is enlarged, in agreement with the
$\phi\sim h^{1/\delta}$ {\it Ansatz} [($\Diamond$) $\varrho=2$,$L=100$,
$\eta=4.2$; ($+$) $\varrho=2$,$L=223$, $\eta=4.2$].  }
\end{figure}

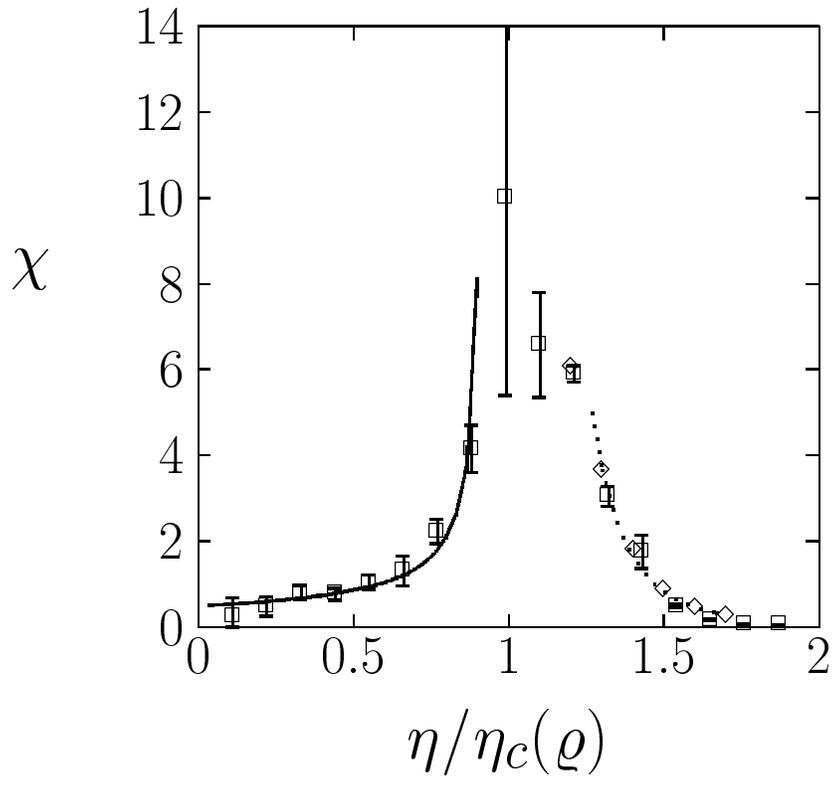
\begin{figure}
\caption{The susceptibility of the system obtained with two different methods:
($\Box$ with error bars) applying definition (15), and ($\Diamond$) from
data collapse in the $\phi_{\mbox{\scriptsize sat}}
>\phi>\phi_{\mbox{\scriptsize noise}}$ regime, which is a more
precise procedure for $\eta>\eta_c$. The data suggest that in this case
the divergence is {\it asymmetric}: $\gamma_\ast'\approx 1$ (solid line)
and $\gamma_\ast\approx 4$ (dotted line).}
\end{figure}

\begin{figure}
\caption{Snapshots of the time development of a system with  $N=4000$, 
$L=40$ and $v_0=0.01$ at $50$ (a), $100$ (b), $400$ (c) and $3000$ (d)
Monte-Carlo steps.  First the behavior is reminiscent of the equilibrium
$XY$ model, where the long range order is missing since vortices are
present in the system.  However the vortices are unstable, and finally a
self-organized long-range order develops.}
\end{figure}


\begin{references}
\bibitem{NoteI} In the uncorrelated case the momentum of the dominant
cluster would be the integral of an uncorrelated noise yielding $\alpha
= 0.5$.
\end{references}
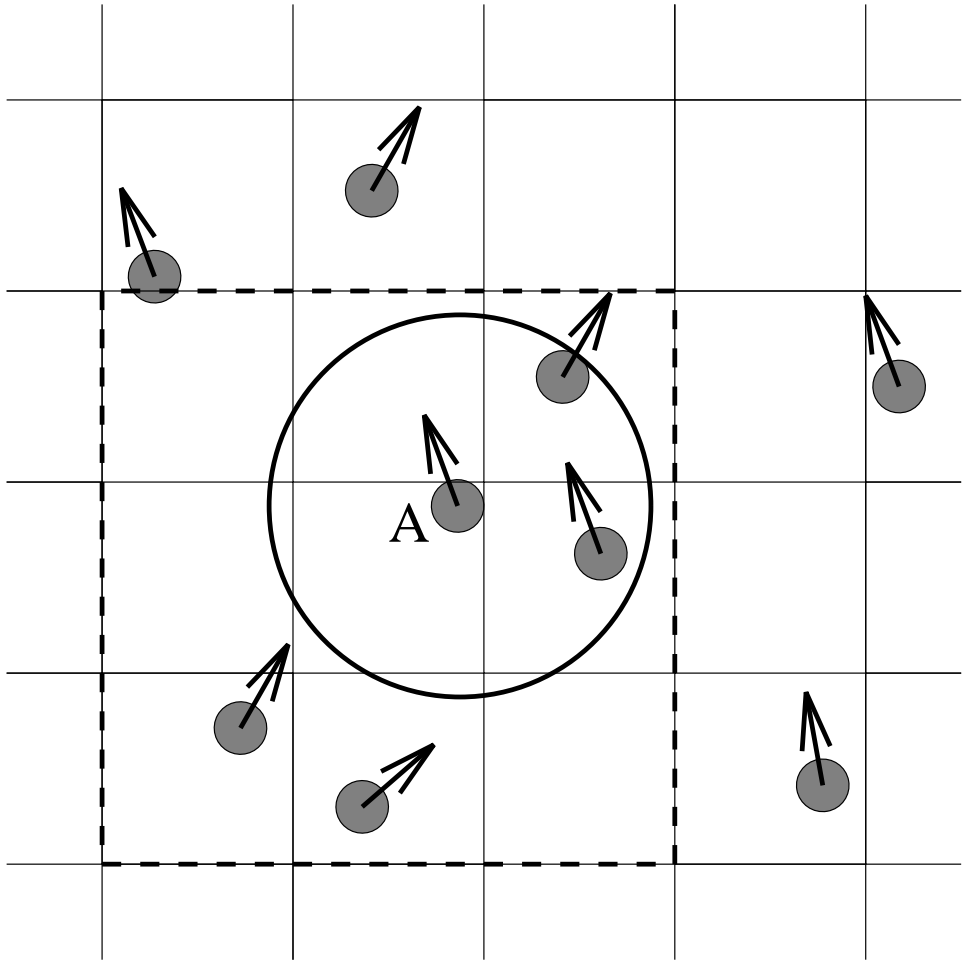
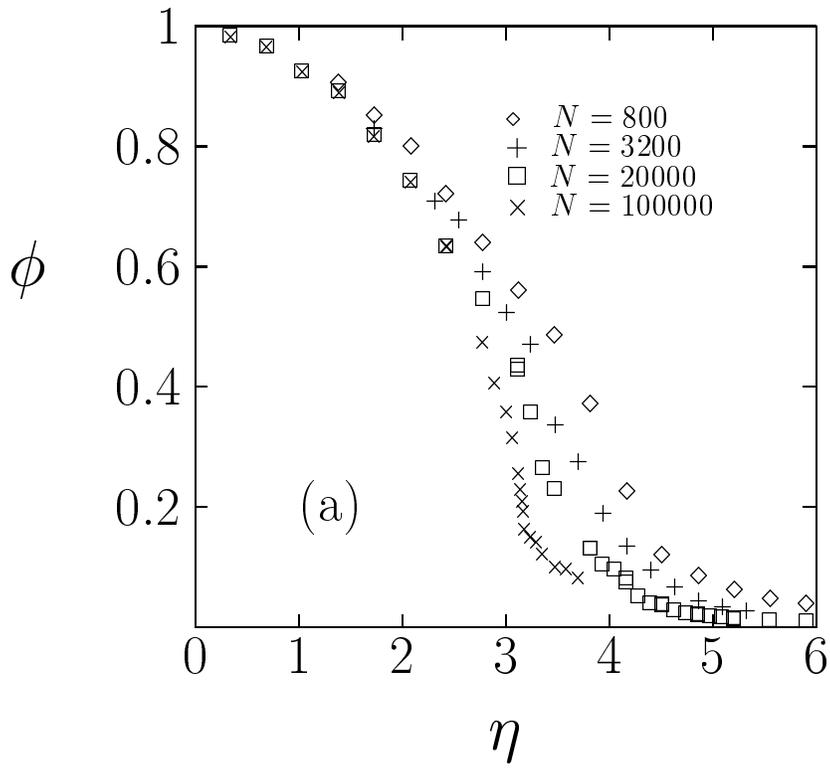
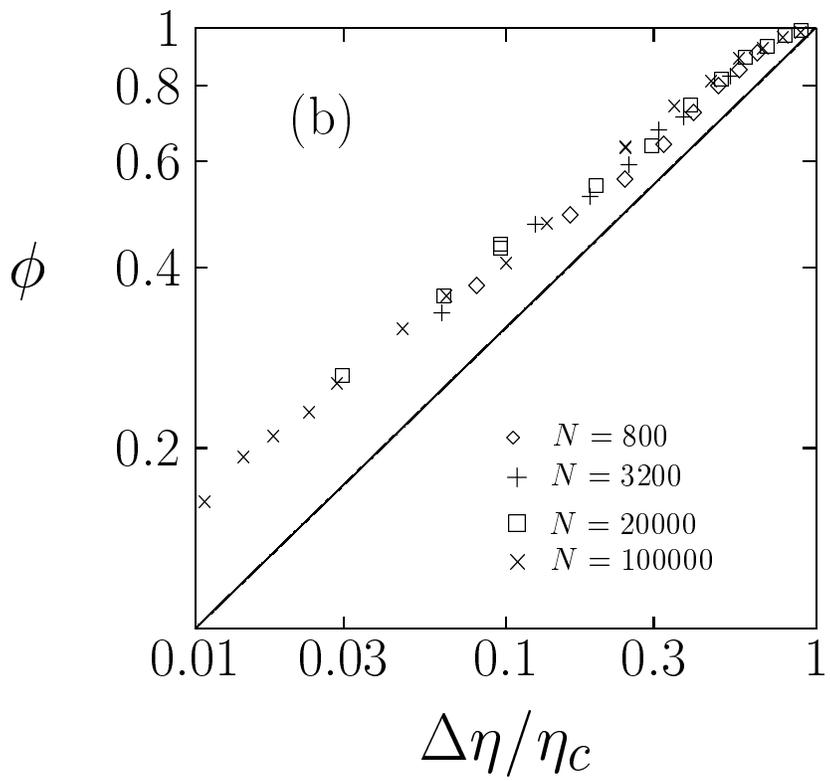
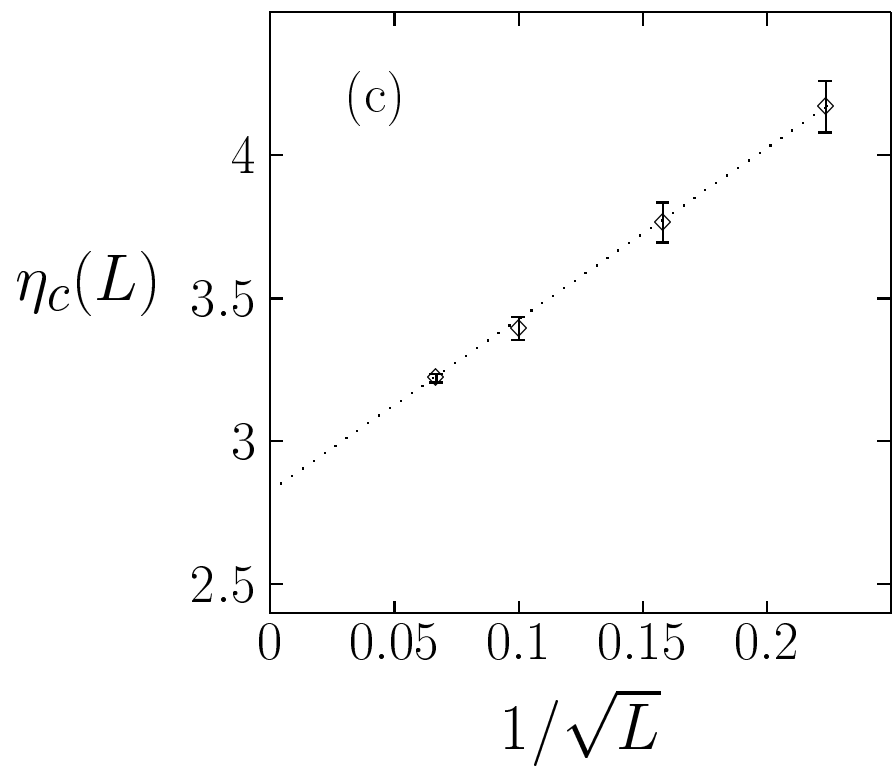
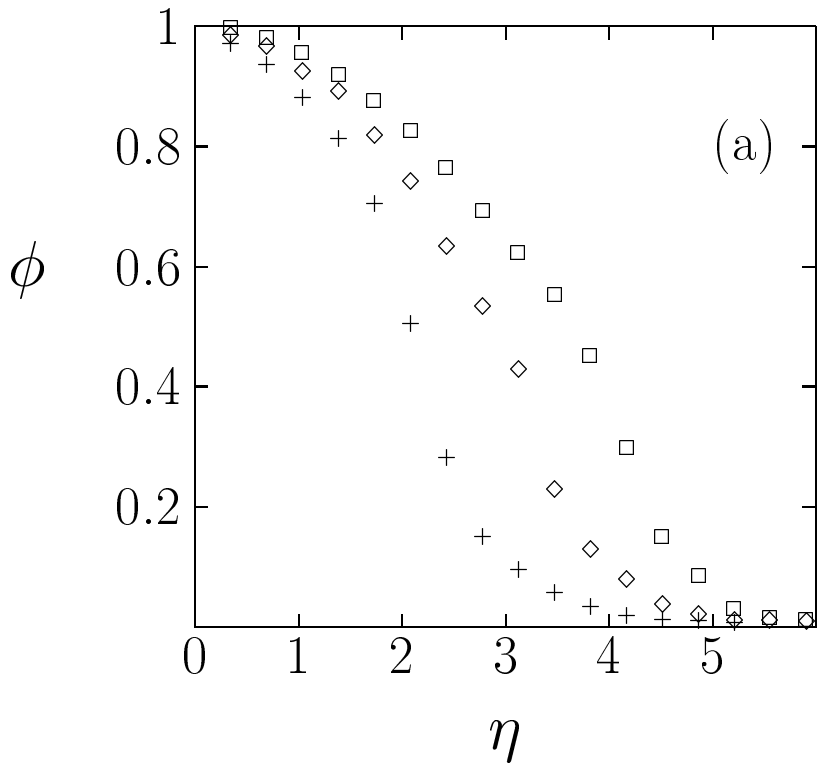
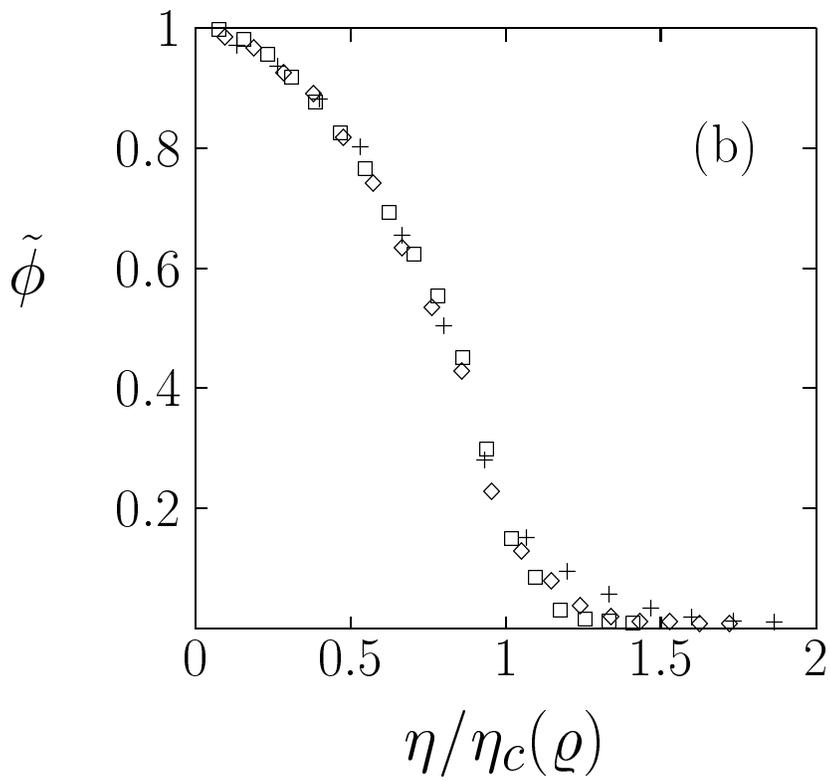
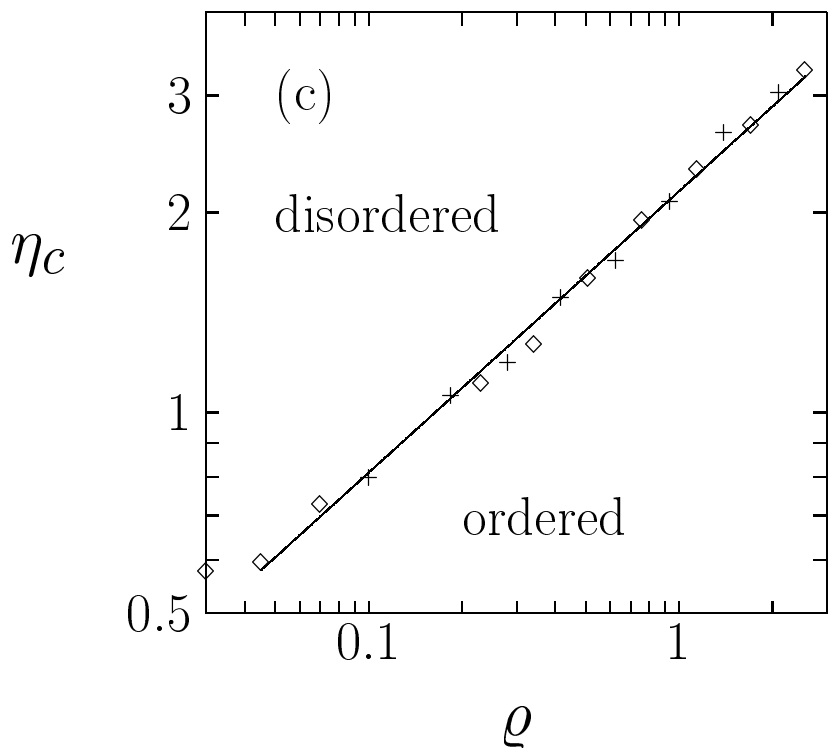
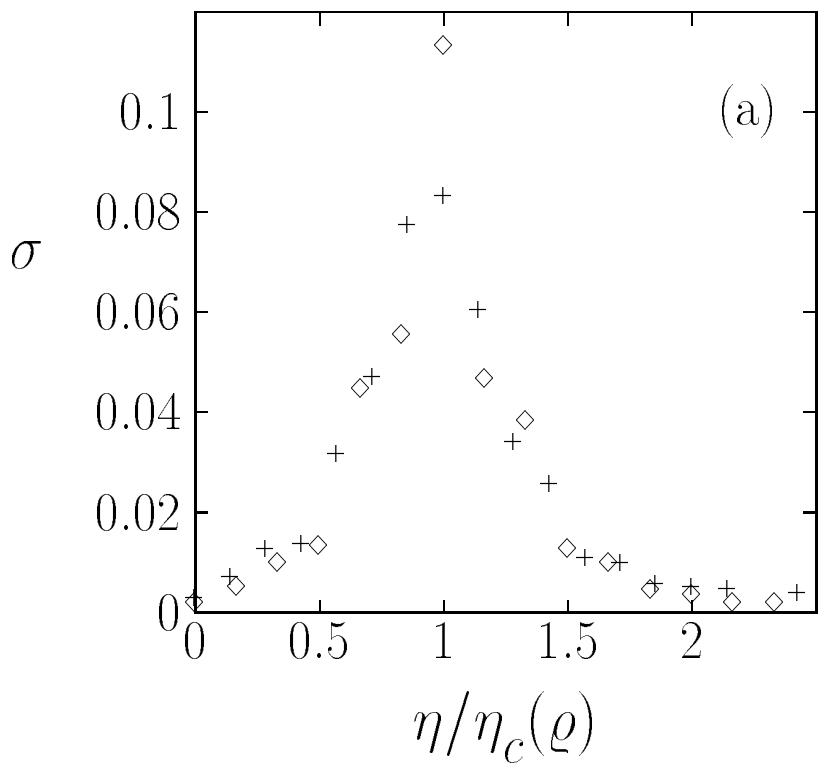
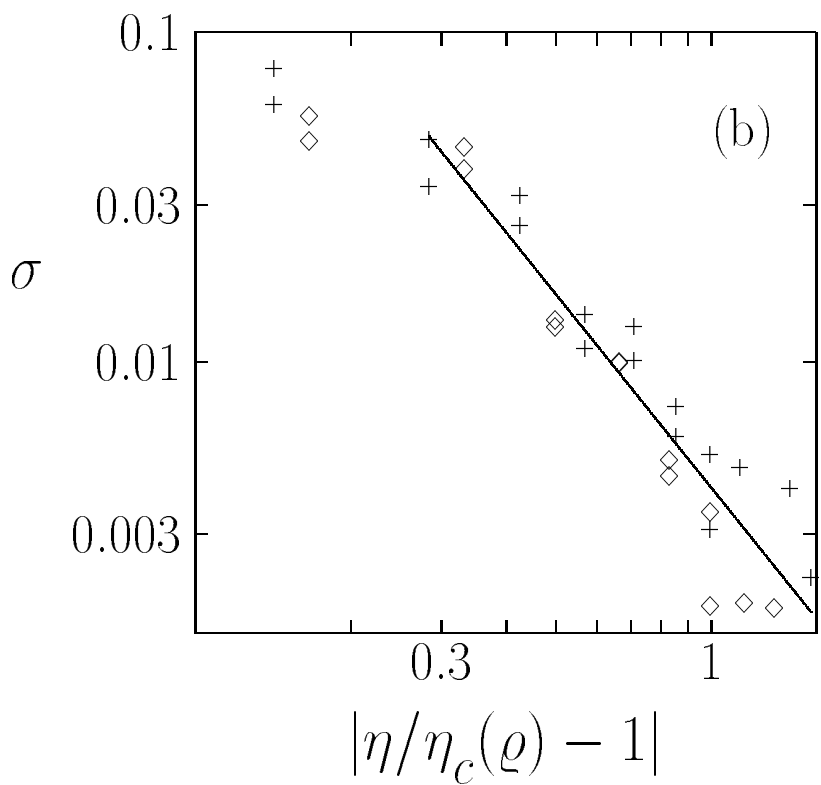
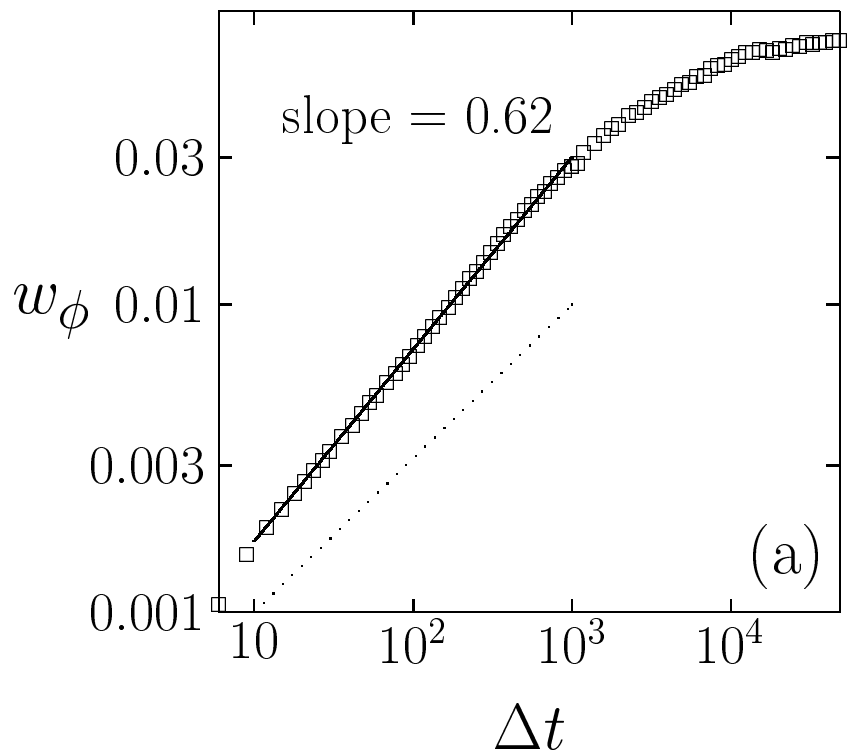
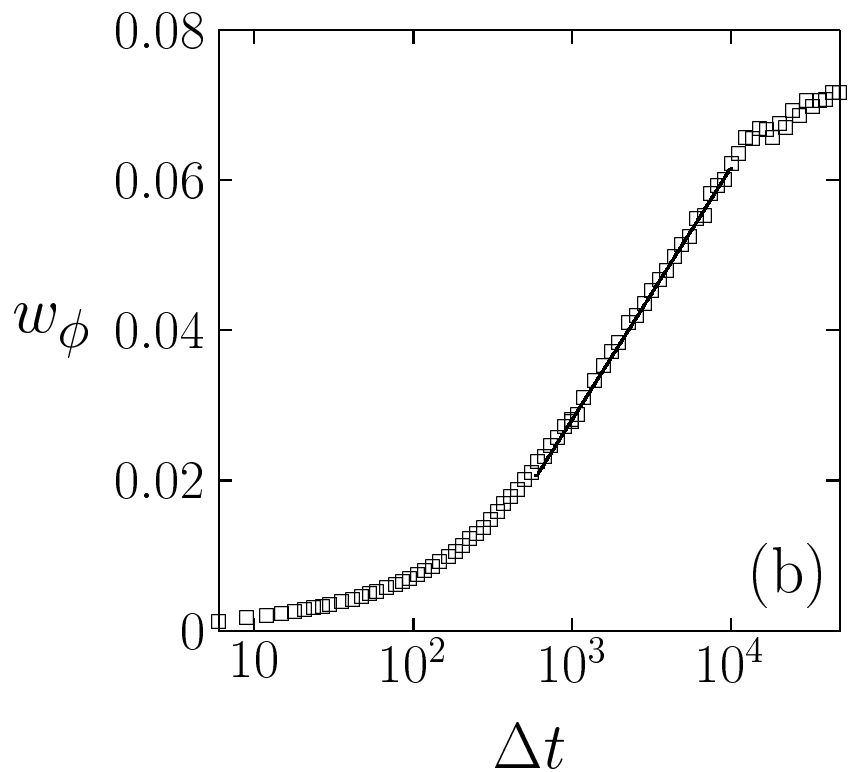
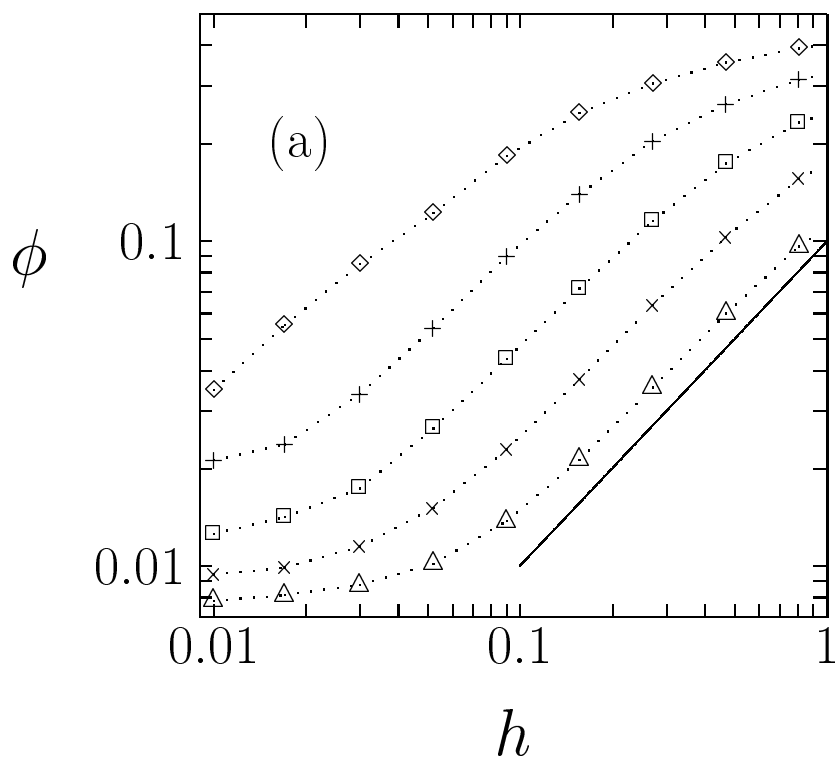
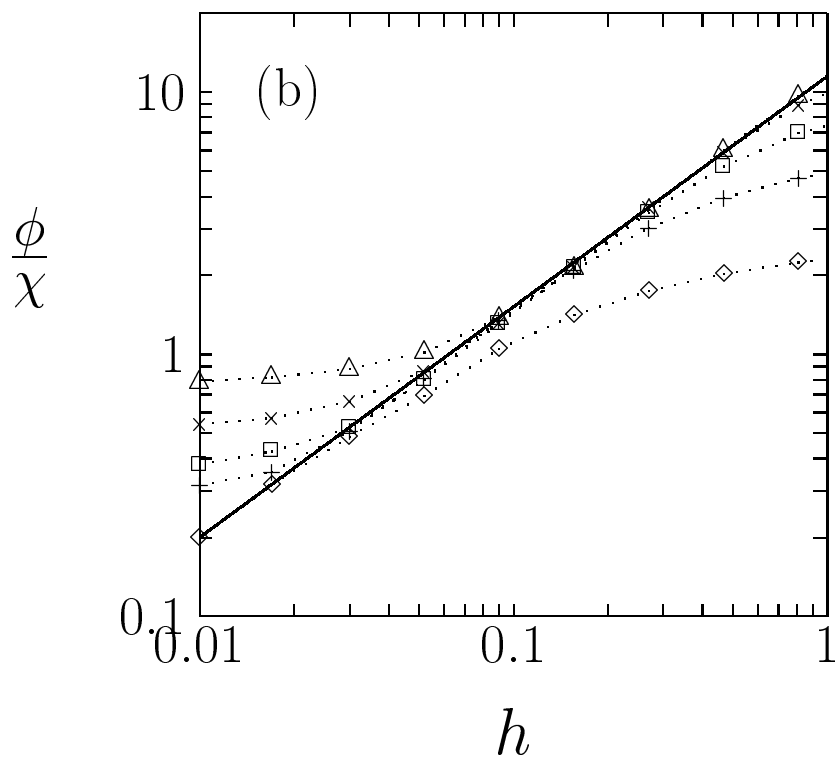
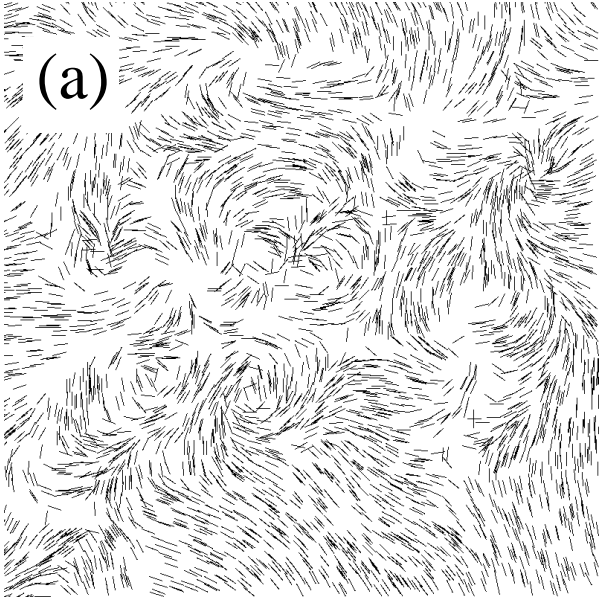
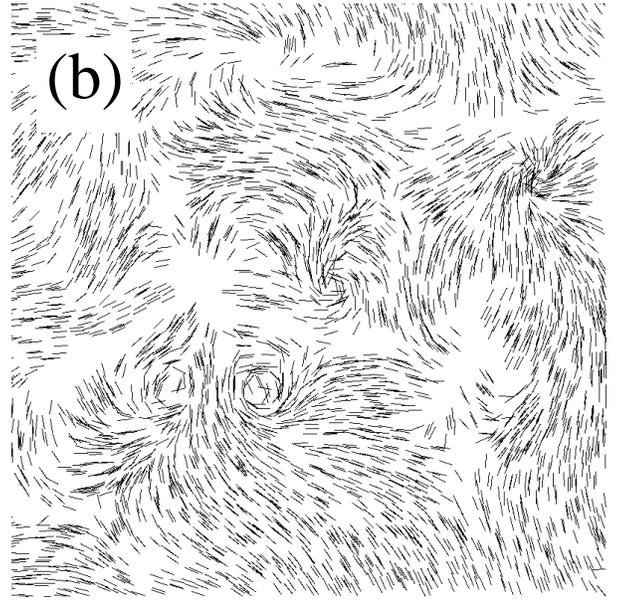
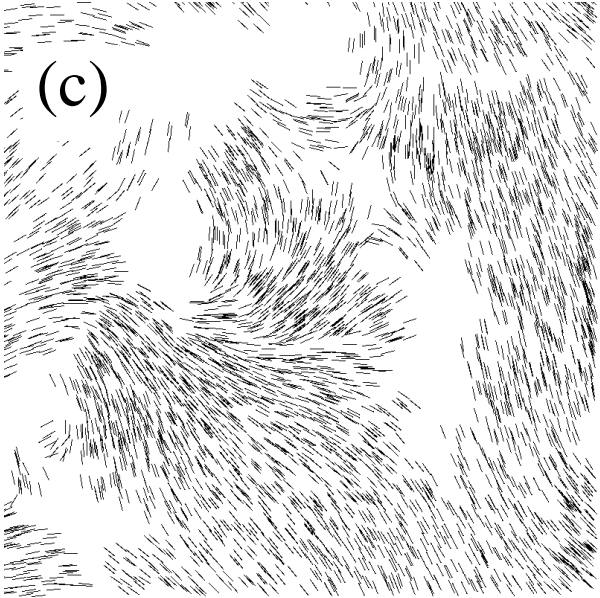
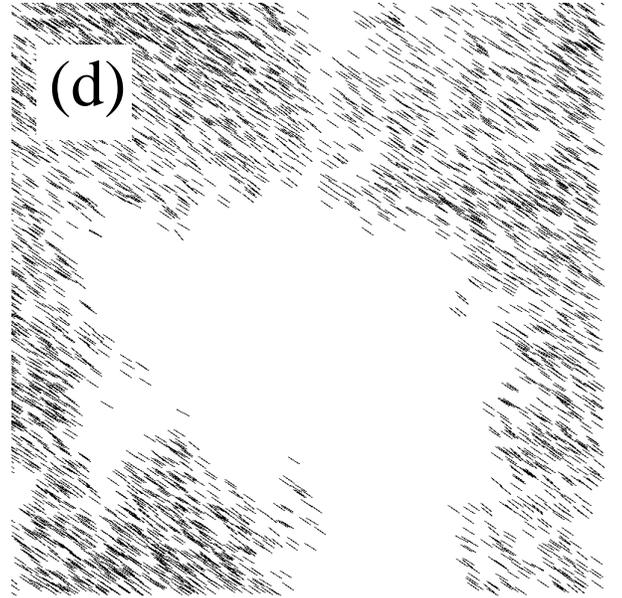
\end{document}